\documentclass[
 twocolumn,
 floatfix,showpacs,preprintnumbers,
 amsmath,amssymb,pre]{revtex4-1}

\usepackage{graphicx}
\usepackage{amssymb}
\usepackage{bm}
\usepackage{xcolor}
\usepackage{pifont}

\renewcommand{\v}[1]{{\boldsymbol #1}}

\DeclareGraphicsExtensions{.pdf}

\begin{document}

\setlength{\voffset}{3\baselineskip}

\title{Control of cellular automata}

\author{Franco Bagnoli}
\email{franco.bagnoli@unifi.it}
\affiliation{Dipartimento di Energetica, Universit\`a di Firenze,
 Via S. Marta 3, I--50139 Firenze, Italy, 
 also CSDC and INFN, sez.\ Firenze.}
\author{Ra\'ul Rechtman}
\email{rrs@cie.unam.mx}
\affiliation{Centro de Investigaci\'on en Energ\'\i a, Universidad Nacional
 Aut\'onoma de M\'exico, Apdo.\ Postal 34, 62580 Temixco, Mor., Mexico.}
\author{Samira El Yacoubi}
\affiliation{Institut de Mod\'elisation et Analyse en G\'eo-Environnement et 
 Sant\'e (IMAGES), Universit\'e de  Perpignan, 52, Paul Alduy Avenue, 
 66860 - Perpignan Cedex. France.}
\email{yacoubi@univ-perp.fr} 

\date{\today}

\begin{abstract}
We study the problem of master-slave synchronization and control of   totalistic cellular automata (CA) by putting a fraction of sites of the   slave equal to those of the master and finding the distance between both as   a function of this fraction. We present three control strategies that
exploit local information about the CA, mainly, the number of nonzero Boolean   derivatives. When no local information is used, we speak of synchronization.   We find the critical properties of control and discuss the best control 
strategy compared with synchronization.
\end{abstract}


\maketitle


\section{Introduction}

There is a certain interest in modelling extended systems using discrete units, in many cases Boolean ones.  There are two main areas for which discrete modelling is appropriate: systems whose dynamics are well approximated by a discrete process, and those for which discrete processes constitute a simplified computational description.  Examples of systems of the first kind are genetic networks~\cite{kaufman}, some formalization of neural networks~\cite{neural}, DNA replication and translation~\cite{DNA}, and VLSI digital circuits. These discrete models are stable and predictable in spite of a noisy environment.  In order to overcome noisy fluctuations affecting signals of electronic devices, a large negative feedback is added, decreasing the overall efficiency but making the system more reliable.  The extreme case of this approach is that of working at the saturation level of devices (e.g., transistors), leading to a digital representation of the problem.  In biology, DNA replication and translation is essentially a digital process, and in many aspects gene control is also a kind of deterministic device, even in the presence of a fluctuating environment.

On the other hand, discrete modelling and in particular cellular automata are widely used as effective computational tools for simulating complex systems~\cite{ACRI}. The idea is that of looking for the emergence of a complex dynamics out of an ensemble of simple components. Some examples are opinion formation, epidemics, and traffic~\cite{Chopard, ComplexCA}.
 
A discrete system is formed by separate units which can be nodes or sites. The state of each unit is taken from a discrete set and evolves in discrete time steps.  Cellular automata (CA)~\cite{CA} are the typical mathematical examples of such systems, even though one may be also interested in non-homogeneous contact networks and non-parallel dynamics~\cite{non-sync}.

We aim to study the dynamical properties of discrete dynamical
systems, and in particular how these characteristics may be exploited
for their control.  In the following we shall consider simple
one-dimensional CA, but most of our considerations can
be extended to more complex systems.

The state of a discrete dynamical system is given by a discrete set of
numbers, and a trajectory is simply given by a list of such
states. Since the total number of possible states is finite, and due
to the deterministic character of the system, after an eventual
transient the trajectory has to enter a cyclic behavior (including
fixed points). However, the size of this periodic attractor and the number of
different attractors may scale with the size of the system. These
numbers may grow exponentially with the system size, and quickly
become intractable for practical purposes. We shall denote these as
discrete extended systems.

Discrete extended systems are not chaotic in the usual sense. They have been
selected or built with the explicit goal of being insensitive to small
perturbations. However, they react to \emph{finite} stimuli (change of
state of some element), as in the experiments of gene knockout or silencing~\cite{genes}. In this case, the reaction may be localized, or
may extend to the whole system. In the latter case, the future
evolution of the system is practically unpredictable. This is a
situation analogous to usual chaotic systems, and the term ``stable
chaos'' has been coined for such systems~\cite{StableChaos}.

Although classical perturbation theory cannot be applied to these
systems, it is possible to define the discrete analogous of
derivatives (Boolean derivatives~\cite{vichniac84,bagnoli92}), as
illustrated in Sec.~\ref{sec:synchro-control}. By using this concept, it is
then possible to construct a Jacobian matrix and to define the largest Lyapunov
exponent, which discriminates well between systems that are ``stable
chaotic'' and others whose evolution is more predictable~\cite{bagnoli99}.

A different, but complementary, approach to the study of the chaotic
properties of a system is based on master-slave 
synchronization~\cite{ReplicaSynchro}. The
idea is that of driving the slave with part of the signal from the
master. It is possible to show that, for simple maps or for a uniform
driving in extended continuous systems, there is a relationship
between the synchronization threshold and the largest Lyapunov
exponent.

A modified version of this synchronization procedure can be applied to
discrete systems. The idea is that of synchronizing them using an all-or-nothing procedure; a fraction of sites are chosen at random and the state in the slave system at those sites, is set equal to that of
of the corresponding sites in the master system (pinching
synchronization)~\cite{bagnoli99}.  Applied to cellular automata, the
pinching synchronization threshold shows a remarkable correlation with
the previously defined largest Lyapunov exponent defined using Boolean
derivatives.

The pinching synchronization procedure defines a directed percolation process among defects, that is, the sites that differ between the master and the slave. We say that the two replicas are not synchronized if in the long time limit  there is some defect surviving  (a more precise definition is given in the following section). 

The synchronized state is absorbing, once it is reached, it cannot be abandoned.  
However, in the absence of a synchronization effort, the synchronized state is unstable  for ``chaotic'' systems, since a finite perturbation will in general
propagate to the rest of the system (as said, this is indeed a possible
definition of a chaotic discrete system). The synchronization efforts
(pinching synchronization) has the goal of making the synchronized
state stable, and the synchronization transition corresponds to the
marginal stability of this state.

By considering now only the points to which the pinching
synchronization is not applied, one has a percolation backbone over
which the defects can survive. Clearly, there is an upper value of the
pinching synchronization probability over which the percolation
backbone disappears and and synchronization takes place regardless of the rule. Differently from 
highly chaotic maps~\cite{MapSynchro}, all cellular automata synchronize well
below such limit, indicating that the self-annihilation of defects
plays a fundamental role in the synchronization transition. Indeed, the disappearance of defects is due both to the pinching procedure (which can be random and uncorrelated) and to the self-annihilation process, which in general is correlated. This correlation has a deep impact on the criticality of the phase transition, as we show in Section~\ref{sec:results}.

We finally arrive to the main point of this paper.  In control theory a controller is used to make a dynamical system behave in a specified way.  The problem of control of a dynamical system may be split into two parts: the driving of the system to a target area in phase space, and the stabilization of a given trajectory originating from this area.  Chaotic systems are ideal targets for control, since their sensitivity to small changes may be exploited to drive them to the target area~\cite{OttDriving}. After that driving phase, chaos may be suppressed in order to make them follow, for instance, a desired periodic orbit~\cite{Ott}.

We shall concentrate on the second part, the stabilization of a given
trajectory. Let us denote with the term ``natural'' a trajectory that is generated by the dynamics of
the system, and with the term ``artificial'' all the other trajectories. In other words, if we initialize the given system in one of the states of a natural trajectory, then the system will follow the
trajectory without any additional external control. On the contrary, ``artificial'' trajectories need a continuous control 
to force the system under study to follow them.

The problem of driving a chaotic system on a natural trajectory
may be seen as the problem of synchronizing a ``slave''  with a
``master'' system that happens to follow the desired trajectory.
However, while in the studies about synchronization one exerts little attention to the optimization of coupling, when formulated as a control problem this becomes a crucial issue.

In synchronization, the effort is applied at every site.  In control
problems, one wants to exploit some knowledge about the system. It is
therefore analogous to a synchronization problem of two different
systems with a ``targeted'' force, that tries to ``kill'' the defects
in an efficient way. We show how the concept of Boolean derivatives
can be exploited to achieve this goal~\cite{bagnoli10}.

The result is however rather surprising: the control efforts should
concentrate on the regions giving birth to a small number of defects,
while chaos can be exploited to ``self-synchronize'' the systems.

In Section~\ref{sec:synchro-control} we present definitions of 
totalistic cellular automata, of Boolean derivatives, and synchronization
and control of CA. Results 
of synchronization and three control strategies
are discussed in Section~\ref{sec:results} and 
in Section~\ref{sec:final} we present some final comments.


\section{Chaos, synchronization and control of cellular automata}
\label{sec:synchro-control}

In this Section we recall the definition of totalistic Boolean cellular automata of range $R$, of Boolean derivatives, and present the
master-slave synchronization and control of cellular automata~\cite{bagnoli92,bagnoli92a,bagnoli99}.

A Boolean cellular automaton of range $R$ is a map $\v{f}:{\cal B}^N\to {\cal   B}^N$ with ${\cal B}=\{0,1\}$, $N$ large, and $\v{f}=(f_0,\dots,f_{N-1})$ with $f_i:{\cal B}^N\to {\cal B}$, $i=0,\dots,N-1$.  A state of the CA is $\v{x}=(x_0,\dots,x_{N-1})$ with $x_i\in {\cal B}$, $i=0,\dots,N-1$. The map $\v{f}$ defines a flow on ${\cal B}^N$,
\begin{equation}
 \label{eq:ca1}
 \v{x}(t+1)=\v{f}(\v{x}(t))
\end{equation}
with $t=0,1,\dots$. In what follows $f_i=f,\,\,\forall i$.

Let ${\cal N}_i$ be a neighborhood of site $i$ with $R$ sites and
\begin{equation}
 s_i=\sum_{j\in{\cal N}_i}x_j,
\end{equation}
with $0\leq s_i\leq R$. For totalistic CA of range  $R$,
\begin{equation}
 x_i(t+1)=f(s_i(t)),
\end{equation}
with $f:\{0,\dots,R\}\to {\cal B}$.  There are $2^{R+1}$ different CA of range $R$, each one defined by a $(R+1)$-tuple $(y_0,\dots,y_R)$ such that $y_j=1$ if the outcome of the CA is one when $s_i=j$ and $y_j=0$ otherwise. In what folows, we use Vichniac's notation $RTC$, with $R$ the range, $T$ is for totalistic, and $C$ the number in base 10 of $(y_0,\dots,y_R)$~\cite{vichniac84}.

The control mechanisms we consider depend on the knowledge of the local expanding properties of the CA given by the Boolean derivatives~\cite{bagnoli92} defined by
\begin{equation*}
 \label{eq:derivative}
 \begin{aligned}
  J_{i,\,j}=&\dfrac{\partial f_i(\v{x})}{\partial x_j}\\
          =&f_i(x_0,\dots,x_j\oplus 1,\dots,x_{N-1})\oplus\\      
           &f_i(x_0,\dots,x_j,\dots,x_{N-1})\\
  &=\begin{cases}
     1 & f_i\: \text{changes when}\: x_j\: \text{changes},\\
     0 & f_i\: \text{does not change when}\: x_j\: \text{changes},
    \end{cases}
 \end{aligned}
\end{equation*}
with $\oplus$ the logical exclusive disjunction or what is the same, the sum
modulo 2. 

Higher order Boolean derivatives can be defined in a similar way and
we can write any function $f$ of range $R$ in the equivalent of a
Taylor or Maclaurin expansion that is exact if the expansion goes up
to $R$-th order derivatives~\cite{bagnoli92}. The exact
MacLaurin expansion of a CA with $R=2$ is
\begin{align}
 f(x,y)=&f(0,0)\oplus%
         \left.\dfrac{\partial f}{\partial x}\right|_{(0,0)}\cdot x\,\oplus\\
        &\left.\dfrac{\partial f}{\partial y}\right|_{(0,0)}\cdot y\oplus%
         \left.\dfrac{\partial^2 f}{\partial x\partial y}\right|_{(0,0)}%
         \cdot x\cdot y
\end{align}
with $\cdot$ the logical conjunction. If the expansion of $f$ contains only the first order Boolean derivatives, we say that $f$ is linear.  As an example, let $R=3$, ${\cal N}_i=\{x_{i-1},x_i,x_{i+1}\}$, and $f(s_i)=1$ if $s_i$ is odd and zero otherwise. This is the $3T10$ CA shown in Table~\ref{tab:lderivative} which is linear since 
\[
 f(x_{i-1}, x_i, x_{i+1})= x_{i-1}\oplus x_i\oplus x_{i+1}.
\] 
In Wolfram's notation, this is rule 150~\cite{wolfram83}. Another example
with $R=3$ is the $3T6$ CA, rule 126 in Wolfram's notation, defined in
such a way that $f(s_i)=1$ if $s_i=1,2$ and zero otherwise shown in
Table~\ref{tab:nlderivative}. It is nonlinear since
\begin{align*}
 f(x_{i-1}, x_i, x_{i+1})=& x_{i-1}\oplus x_i\oplus x_{i+1}\oplus \\
                        & x_{i-1}x_i \oplus x_{i-1}x_{i+1}\oplus x_i x_{i+1}.
\end{align*}
\begin{table}
 \caption{\label{tab:lderivative} Truth table and Boolean derivatives
  of the $3T10$ CA.}
   \begin{tabular}{|ccc|c|c|c|c|c|c|c|c|c|}
    \hline
    $x_{i-1}$ & $x_i$ & $x_{i+1}$ & $s_i$& $f$ 
    & $\partial f/\partial x_{i-1}$
    & $\partial f/\partial x_i$
    & $\partial f/\partial x_{i+1}$        \\[1mm]
    \hline                                             
    0    & 0 & 0 & 0 & 0 & 1 & 1 & 1          \\
    0    & 0 & 1 & 1 & 1 & 1 & 1 & 1          \\
    0    & 1 & 0 & 1 & 1 & 1 & 1 & 1          \\
    0    & 1 & 1 & 2 & 0 & 1 & 1 & 1          \\
    1    & 0 & 0 & 1 & 1 & 1 & 1 & 1          \\
    1    & 0 & 1 & 2 & 0 & 1 & 1 & 1          \\
    1    & 1 & 0 & 2 & 0 & 1 & 1 & 1          \\
    1    & 1 & 1 & 3 & 1 & 1 & 1 & 1          \\
    \hline
   \end{tabular}
\end{table}
\begin{table}
 \caption{\label{tab:nlderivative} Truth table and Boolean derivatives
  of the $3T6$ CA.}
   \begin{tabular}{|ccc|c|c|c|c|c|c|c|c|c|}
    \hline
    $x_{i-1}$ & $x_i$ & $x_{i+1}$ & $s_i$& $f$ 
    & $\partial f/\partial x_{i-1}$
    & $\partial f/\partial x_i$
    & $\partial f/\partial x_{i+1}$        \\[1mm]
    \hline                                             
    0    & 0 & 0 & 0 & 0 & 1 & 1 & 1          \\
    0    & 0 & 1 & 1 & 1 & 0 & 0 & 1          \\
    0    & 1 & 0 & 1 & 1 & 0 & 1 & 0          \\
    0    & 1 & 1 & 2 & 1 & 1 & 0 & 0          \\
    1    & 0 & 0 & 1 & 1 & 1 & 0 & 0          \\
    1    & 0 & 1 & 2 & 1 & 0 & 1 & 0          \\
    1    & 1 & 0 & 2 & 1 & 0 & 0 & 1          \\
    1    & 1 & 1 & 3 & 0 & 1 & 1 & 1          \\
    \hline
   \end{tabular}
\end{table}

\begin{figure}[t]
 \begin{tabular}{cc}
  (a) & (b) \\
  \includegraphics[width=0.5\columnwidth]{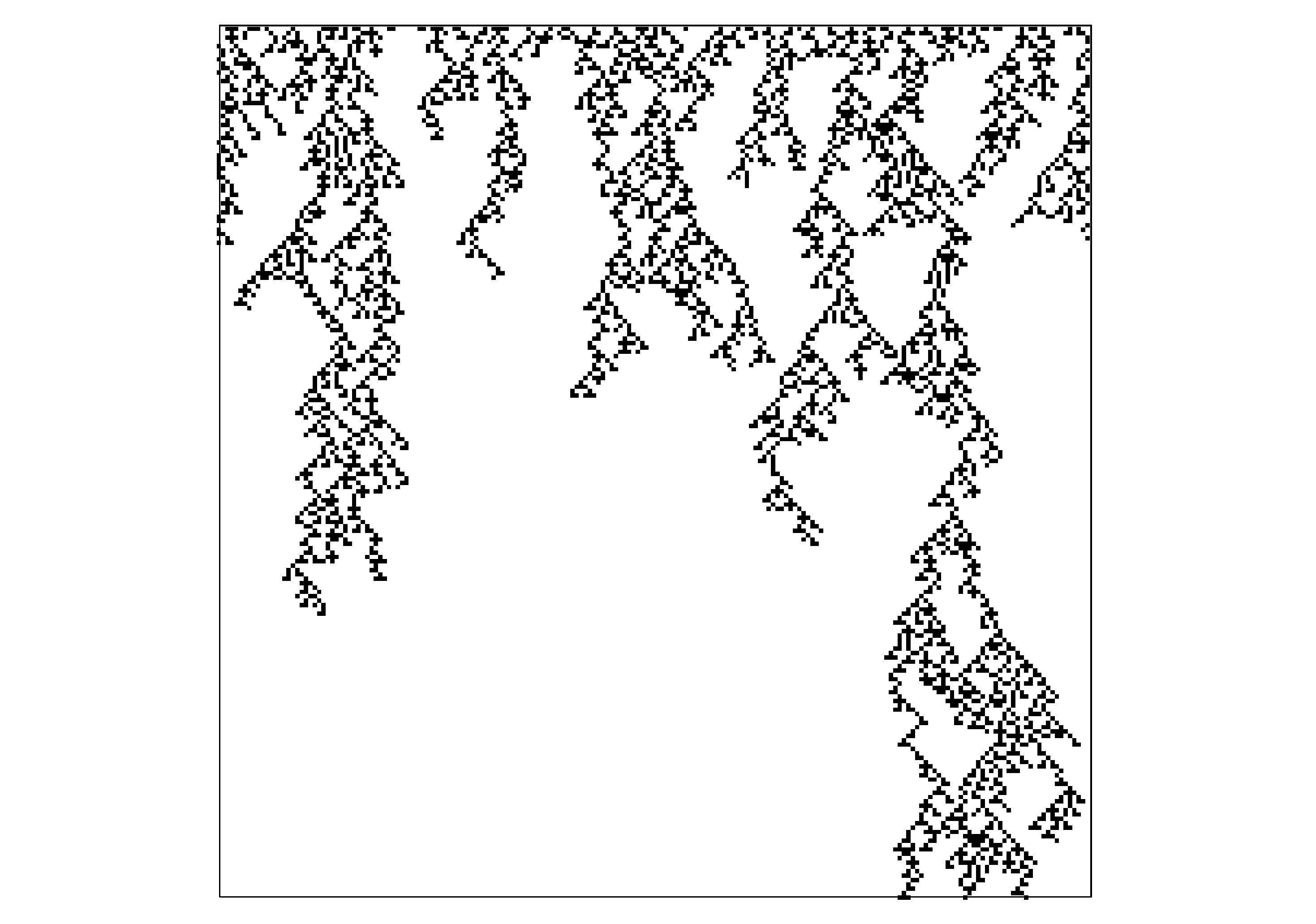} &
  \includegraphics[width=0.5\columnwidth]{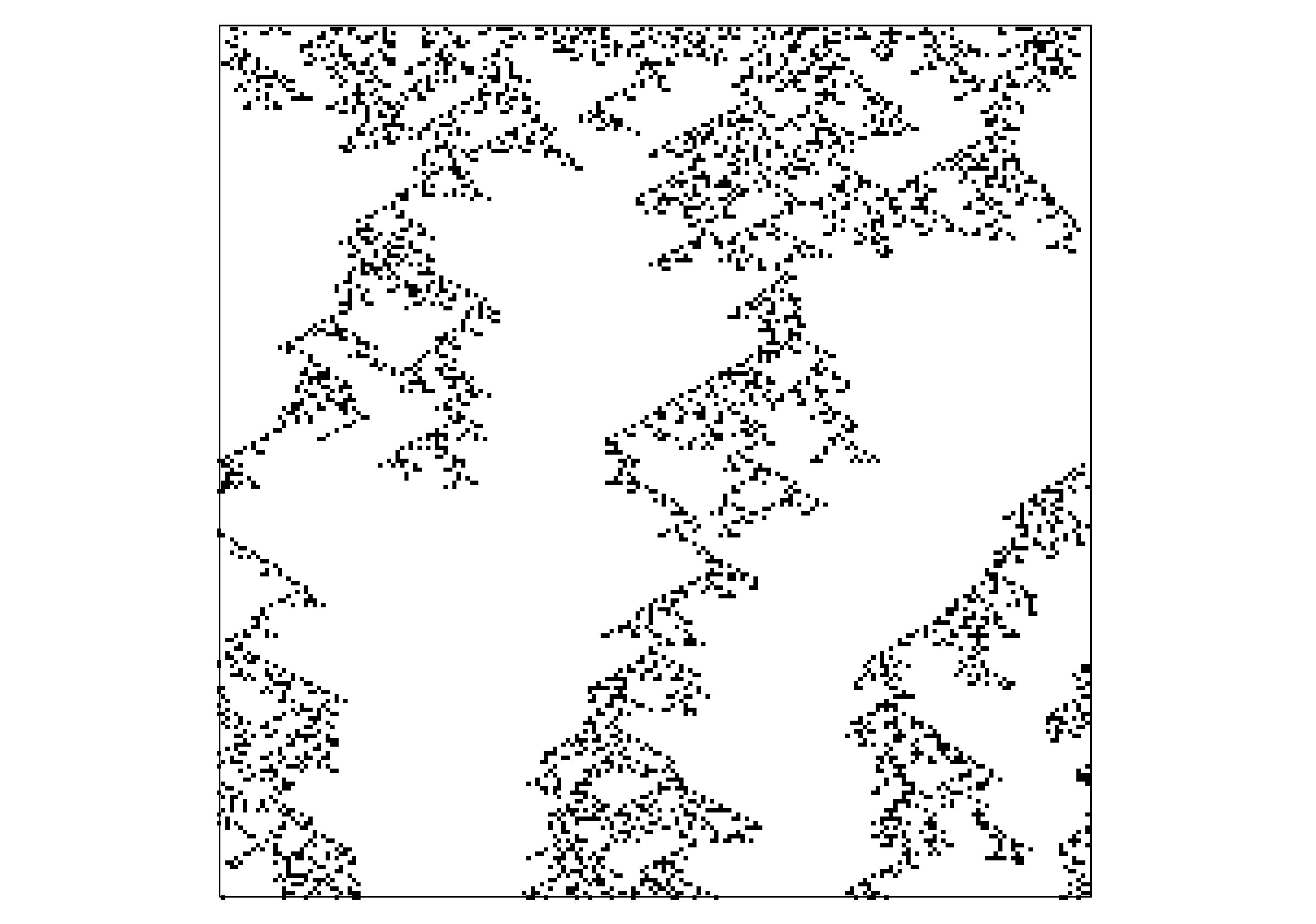}
 \end{tabular}
 \caption{\label{fig:perc} Time evolution of the difference $\v{h}$
   for (a) CA $3T10$, (b) CA $5T42$ with $N=200$ during 200 time
   steps. Time runs from top to bottom.}
\end{figure}

We now discuss a master-slave synchronization and control mechanism of cellular automata where the configuration $\v{x}$ represents the master and $\v{y}$ the slave. Let $p_i$, $i=0,\dots,N-1$ be a Boolean variable that is 1 with probability $\pi_i$ and 0 otherwise. That is,
\[
 p_i= 
 \begin{cases}
  1 & \text{with probability}\: \pi_i\\
  0 & \text{with probability}\: 1-\pi_i.
 \end{cases}
\]
Then 
\begin{align*}
 x_i(t+1)&=f(\v{x}(t)),\\
 y_i(t+1)&=(1-p_i(t))\cdot f(\v{y}(t))\oplus p_i(t)\cdot%
             f(\v{x}(t)),\\
   &=\begin{cases}
      f(\v{x}(t)) & \text{if $p_i(t)=1$}\\
      f(\v{y}(t)) & \text{if $p_i(t)=0$},\\
     \end{cases}\\ 
  h_i(t+1)&=y_i(t+1)\oplus x_i(t+1).
\end{align*}
In the last Eq. $h_i=1$ if $y_i$ is different from $x_i$ and is zero otherwise. We use the shorthand $\v{h} = \v{y} \oplus \v{x}$.  Wherever $p_i=1$, $y_i=x_i$ and we say that $\v{y}$ and $\v{x}$ are ``pinched'' at site $i$.  We define the control parameter $k$ and the order parameter $h$ by
\begin{equation}
 k=\dfrac{1}{N}\sum_i p_i,\qquad h=\dfrac{1}{N}\sum_i h_i.
\end{equation}
If $k=0$, $\v{y}$ evolves independently of $\v{x}$ and $h\neq 0$, even though the self-annihilation of defects is still present due to the Boolean character of the system. If $k=1$, $h=0$, both states are pinched together. There is a threshold $k_c$ above which $h=0$ in the long time limit.  See Fig.~\ref{fig:perc} for a visual illustration of the synchronization process.
If no knowledge is assumed about the local behavior of $f$ that
defines the CA, we speak of pinching synchronization~\cite{bagnoli99} and take
$\pi_i=p$ for all $i$. In the case where some knowledge about $f$ is
used to determine $\pi_i$ (that can change in time), we speak of pinching
control.
\begin{figure}[t]
 \includegraphics[width=\columnwidth]{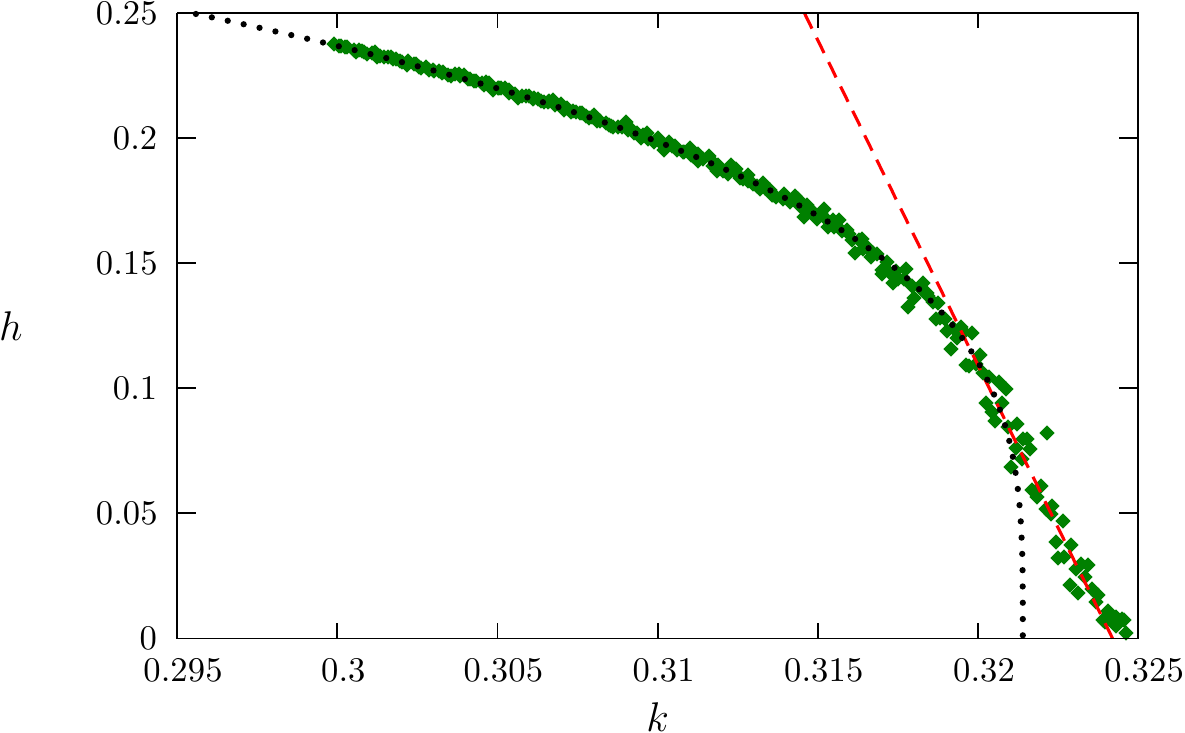}
 \begin{picture}(0,0)
  \put(-80,40){\includegraphics[width=0.45\columnwidth]{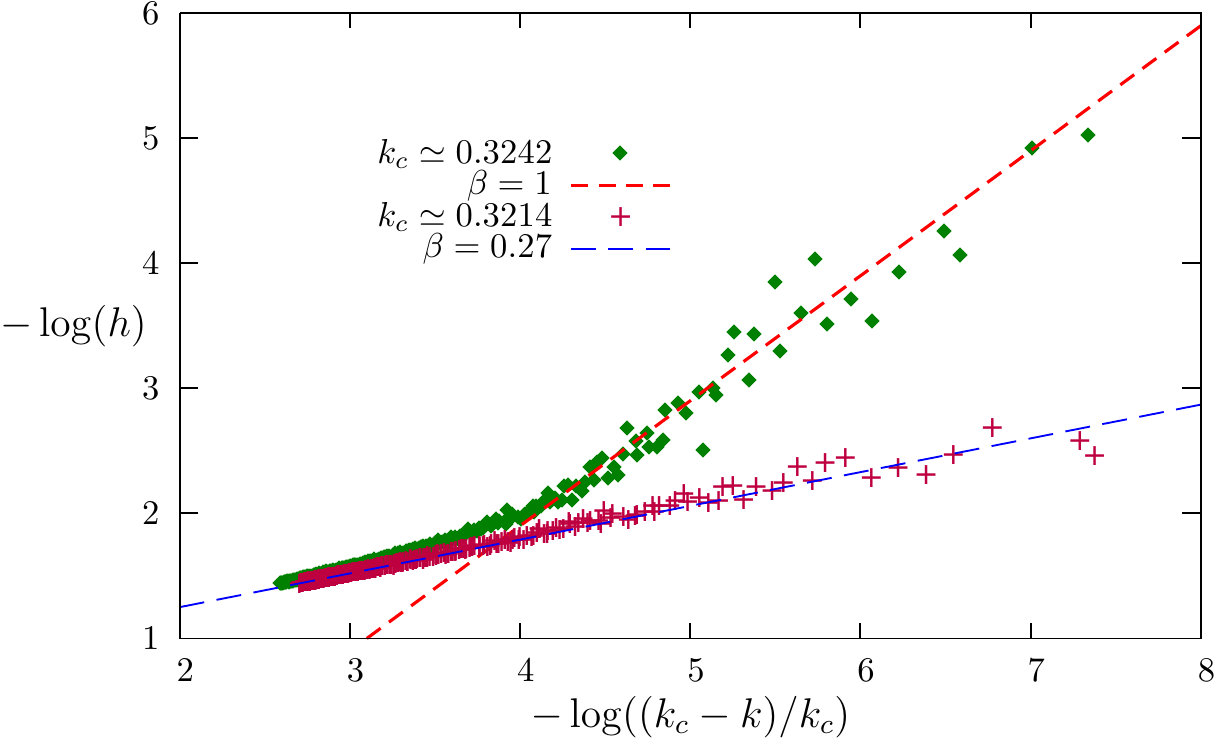}}
 \end{picture}
 \caption{\label{fig:3T10}  The scaling of the asymptotic distance $h$
   as a function of the synchronization effort $k$ for rule 3T10 in
   case {\bf 0} and $N=20,000$. The curves follow the scaling law
     Eq.~(\ref{eq:scaling-hk}).  The dotted  curve corresponds to 
     $\beta= 0.27$ ($k_c\simeq 0.321$), compatible with
     directed percolation  and the dashed line
     corresponds to $\beta = 1$   ($k_c\simeq 0.3275$) compatible with a mean field  behavior. In
     the inset, the scaling behavior is reported on a log-log
     scale. }
 \end{figure}

\begin{figure}[t]
 \includegraphics[width=\columnwidth]{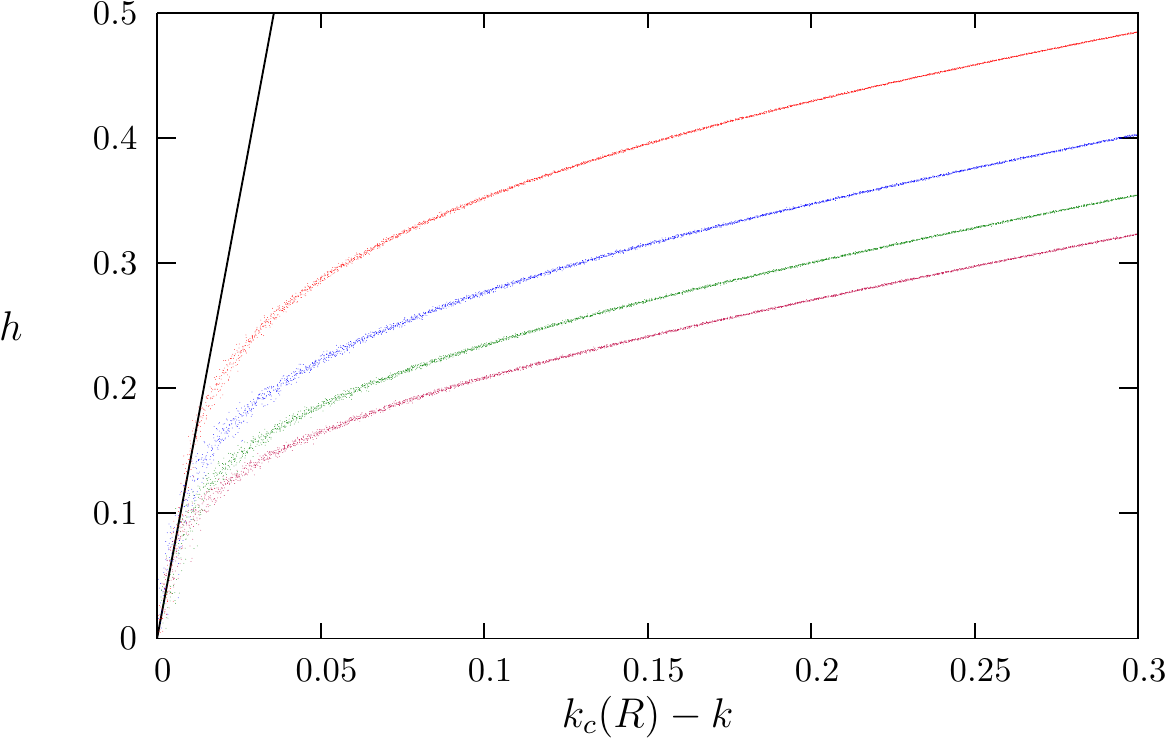}
 \caption{\label{fig:linear} Synchronization 
  (control type 0) of some linear cellular automata, from top
  to bottom: $3T10$, $5T42$, $7T170$, $9T682$, rescaled to make the
  critical point coincide with the origin of axis. In these simulations
  $N=4,000$.  }
\end{figure}

In the following Sec. we show results for synchronization and three 
types of control based on the local expanding properties of the map $f$.
We define $J_i$ as the sum of the $R$ Boolean derivatives that are affected by  site $i$
\begin{equation}
 \label{eq:Ji}
 J_i=\sum_{j\in{\cal N}_i} J_{j,\,i},
\end{equation}
with $0\leq J_i\leq R$ (notice the indices).
The cases we consider are
\begin{itemize}
 \item[{\bf 0}.] $\pi_i(t)=p$ with $0\leq p\leq 1$ independently of
  $i$ and $t$.
 \item[{\bf 1}.] $\pi_i(t)$ is  proportional to $J_i(t)$.
 \item[{\bf 2}.] If $J_i(t)=0$, $\pi_i(t)=0$, otherwise  $\pi_i(t)$ is    
  proportional to $R-J_i(t)+1$.
 \item[{\bf 3}.] If $J_i(t)=0$, $\pi_i(t)=0$, otherwise $\pi_i(t)$ is    
  proportional to $(R-J_i(t)+1)^2$.
\end{itemize}
Case {\bf 0} corresponds to synchronization, no knowledge on the 
dynamics of the CA is used. The other cases, use the derivatives in one
way or the other. Since $0\leq J_i\leq R$, case {\bf 1} will favor
that states $\v{x}$ and $\v{y}$ be pinched at site
$i$ at time $t$ when $J_i(t)$ is large. On the other hand, cases {\bf 2}
and {\bf 3} favor pinching at site $i$ at time $t$ whenever $J_i(t)$ 
is small (case {\bf 3} more than case {\bf 2}).


\section{Results}
\label{sec:results}

\begin{figure}
 \includegraphics[width=\columnwidth]{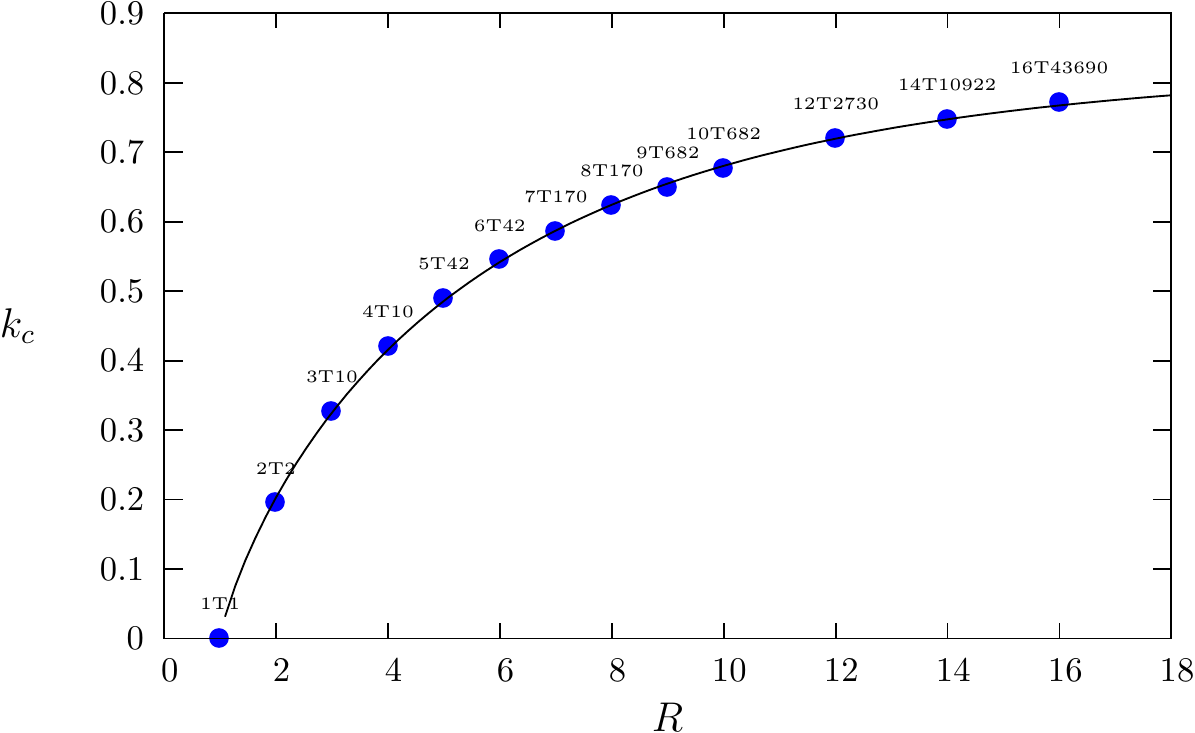}
 \caption{\label{fig:kc} The relation between $k_c$, and  the range $R$ for
  linear  cellular automata, control type \textbf{0}. The solid line marks a
  the best fit of Eq.~(\ref{eq:ks}) with
  $k_\infty=0.825$, $\gamma=0.282$, and $\alpha=0.828$.} 
\end{figure}
\begin{figure}[t]
 \includegraphics[width=\columnwidth]{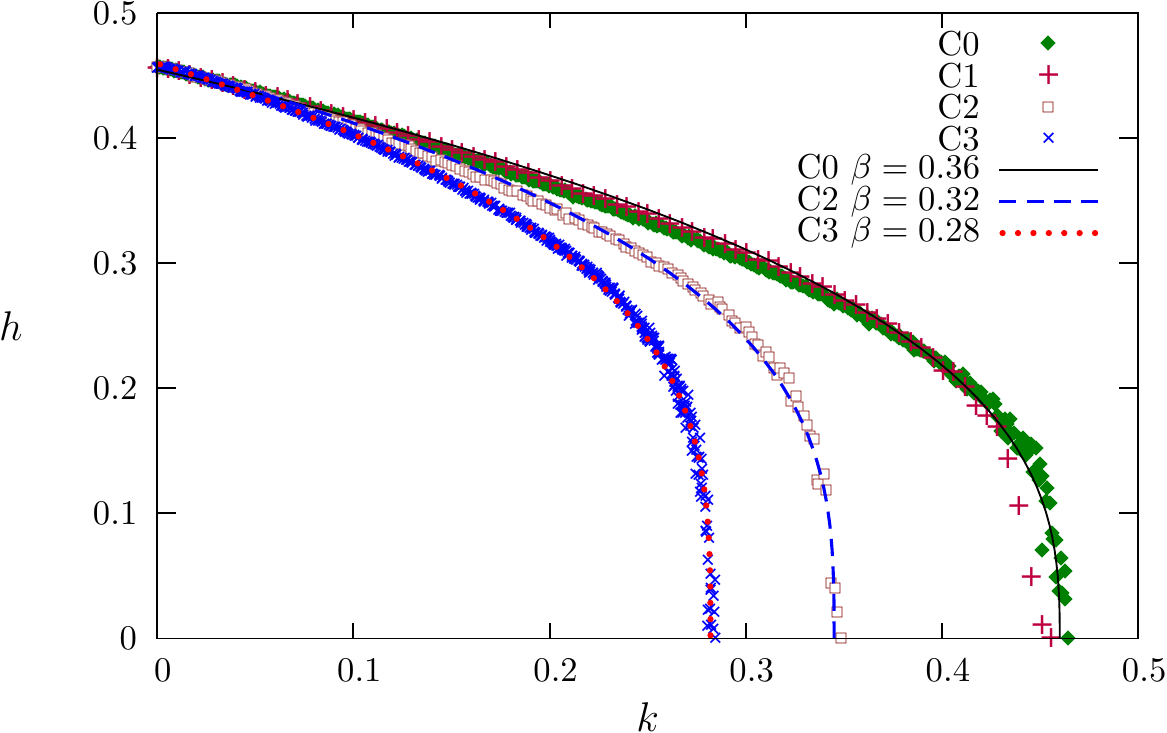}
 \caption{\label{fig:7T10} Critical behavior of nonlinear CA 7T10 for control 
  types ${\bf 0},\dots,{\bf3}$.  In all cases Eq.~(\ref{eq:scaling-hk}) fits
  the data with $h_0\simeq 0.46$ and different values of $k_c$ and $\beta$
  (continuous curves).  Control type \textbf{0} and \textbf{1} agree  
  numerically with $\beta\simeq 0.36$, control type \textbf{2} with
  $\beta=0.32$ control type \textbf{3} with $\beta=0.28$. }
\end{figure}

We begin by presenting the synchronization and control of linear CA.
All types of control coincide with the synchronization case {\bf 0}
for linear CA, since $J_i=R$ for every site $i$.  

Near $k_c$, $h$ follows the scaling law
\begin{equation}
 \label{eq:scaling-hk}
 h=h_0\left(\frac{k_c-k}{k_c}\right)^\beta,
\end{equation}
with $h_0$ the average number of active sites for the unperturbed
rule ($k=0$). 

In Fig.~\ref{fig:3T10} we show the critical behavior of $h$ for CA $3T10$ that has a crossover between the directed percolation (DP, $\beta\simeq0.27$) and the mean field (MF, $\beta=1$) universality classes.  As we show in Fig.~\ref{fig:linear}, other linear rules exhibit the same type of crossover from the MF to the DP universality class. As $R$ grows, the MF character of the transition occupies a smaller region.  This behavior is rather unexpected, since the Grassberger-Jensen conjecture~\cite{conjecture} should hold in this case. The interactions are short-range and, since the visual behavior is irregular, correlations are expected to be short-range.  By applying a transformation $y=1\oplus x$ for odd neighbourhoods one can show that for a linear CA, $h_0=1/2$.

In Fig.~\ref{fig:kc} we show the synchronization threshold $k_c$ of linear totalistic CAs as a function of the range $R$. The numerical data are well fitted by 
\begin{equation}
 \label{eq:kc}
  k_c = k_\infty\left(1-\exp\left(-\gamma(R-1)^\alpha\right)\right).
\end{equation}
We assumed that for $R=1$ the critical value of $k_c$ is zero. In this case each site only has one neighbor,  and therefore any infinitesimal control is able to synchronize the replicas in the long time limit.
When $R\rightarrow\infty$, $k_\infty\simeq 0.825$.

For the nonlinear CA $7T10$, the scaling behavior near the transition is more similar to a DP character, for all types of control, as shown in Fig.~\ref{fig:7T10}, even though the measured exponents differ numerically from the DP exponent. 

For linear CA, all control cases are equivalent. This is not the case for nonlinear CA. Although we have not performed an exhaustive search, some nonlinear chaotic CA show the general pattern $k_c^{(3)}\leq k_c^{(2)}\leq k_c^{(0)}\leq k_c^{(1)}$ with $k_c^{(n)}$ the critical value of the control parameter $k$ in case $n$, $n={\bf 0},\dots,{\bf 3}$. Examples are
CA $7T6$, $7T30$, and $3T6$ whose critical behavior is shown in the top
three graphs of Fig.~\ref{fig:7T10}. For the CA of the second line
of Fig.~\ref{fig:7T10} control {\bf 3} shows the smallest $k_c$.
In the last line of Fig.~\ref{fig:7T10} we show three nonlinear CA for
for which the four types of control are similar.

\begin{figure}[t]
 \begin{tabular}{ccc}
  7T6 & 7T30 & 3T6\\
   \includegraphics[width=0.32\columnwidth]{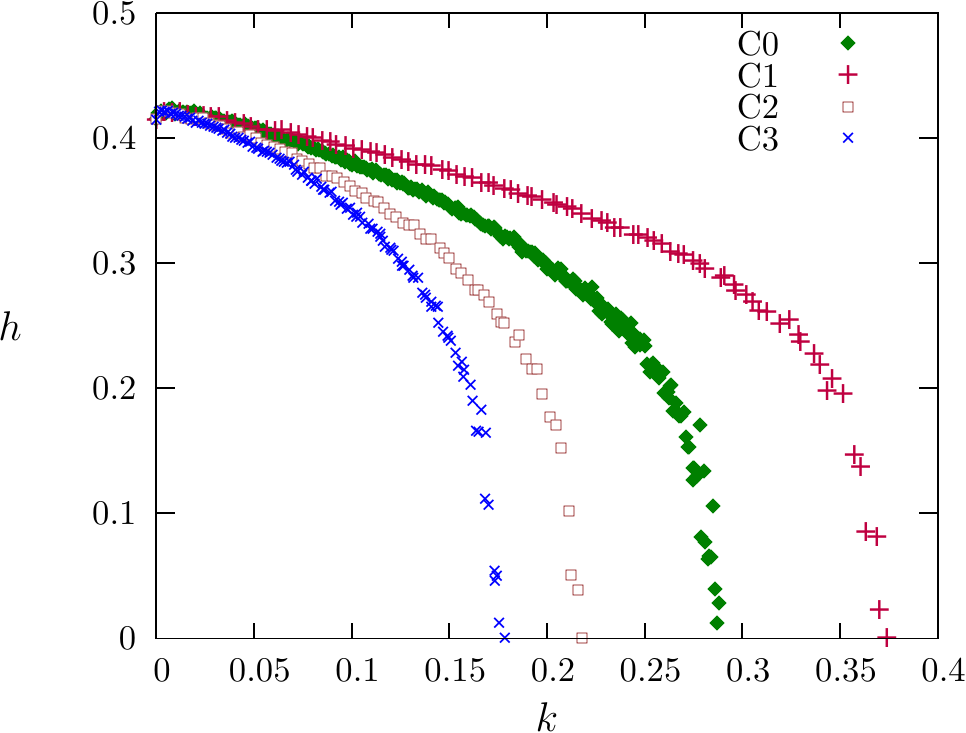} &
  \includegraphics[width=0.32\columnwidth]{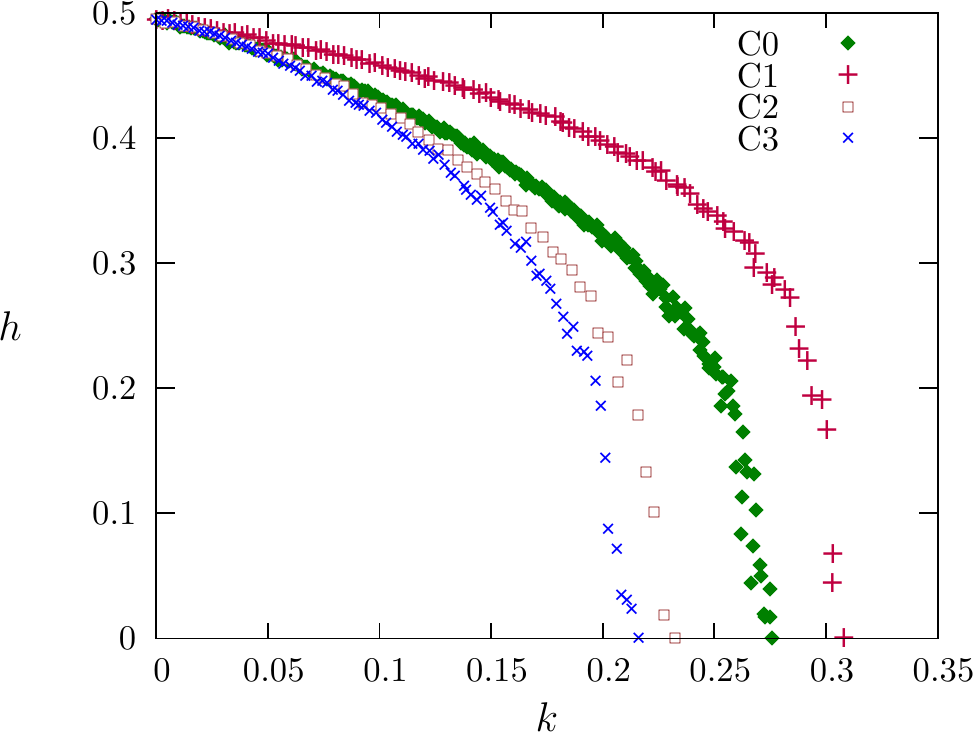} &
  \includegraphics[width=0.32\columnwidth]{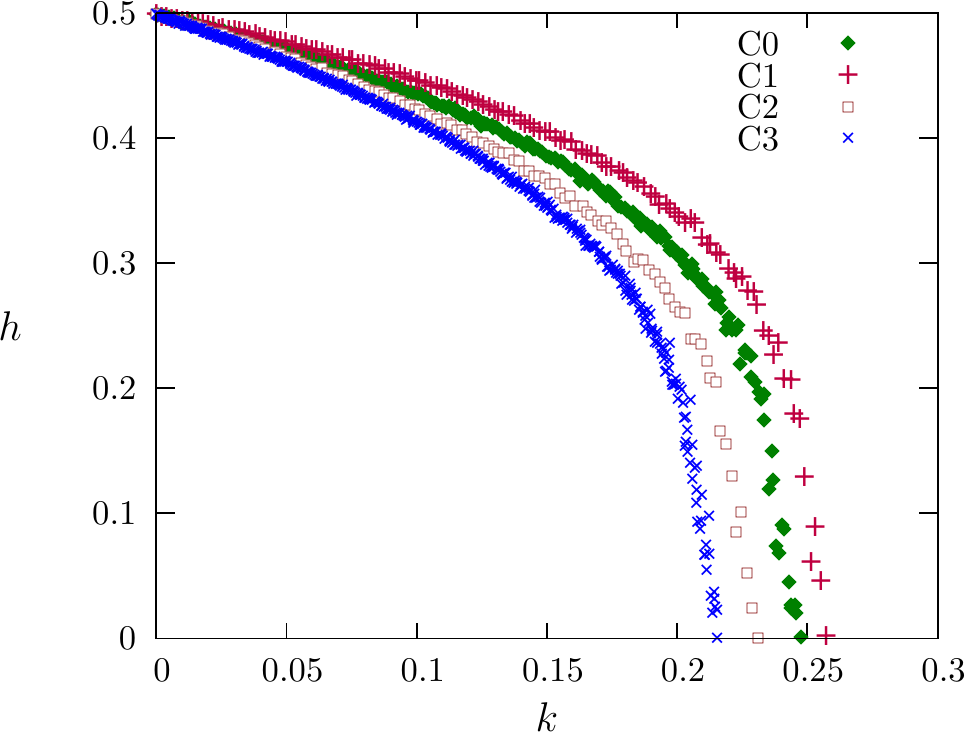}\\
    6T60 & 5T44 &  5T46\\
  \includegraphics[width=0.32\columnwidth]{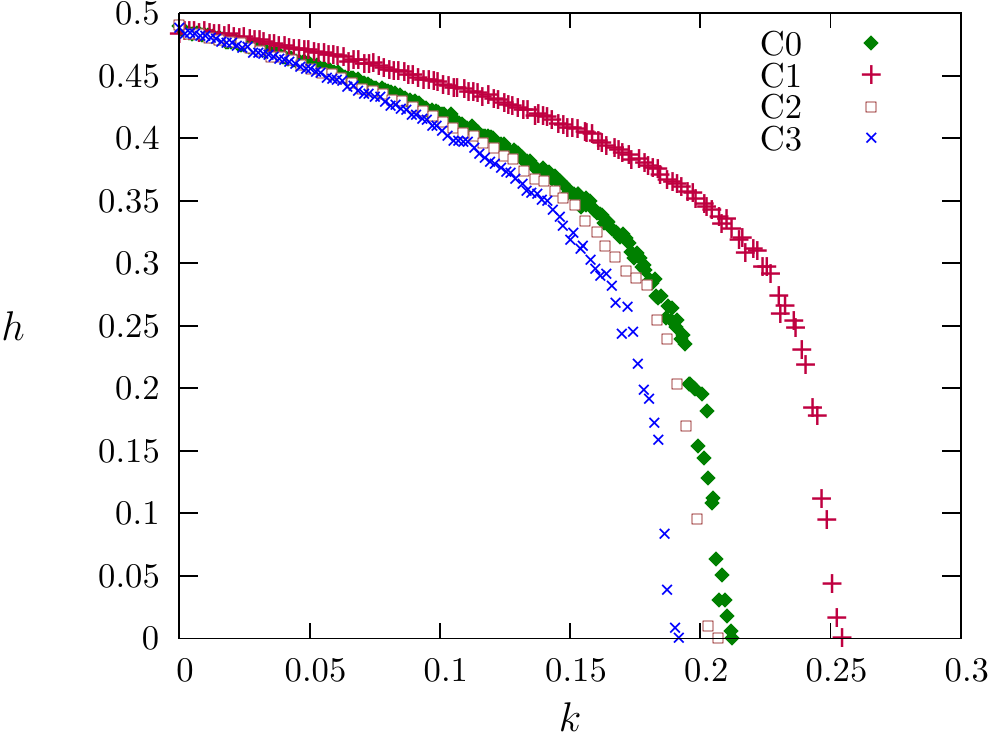} &
  \includegraphics[width=0.32\columnwidth]{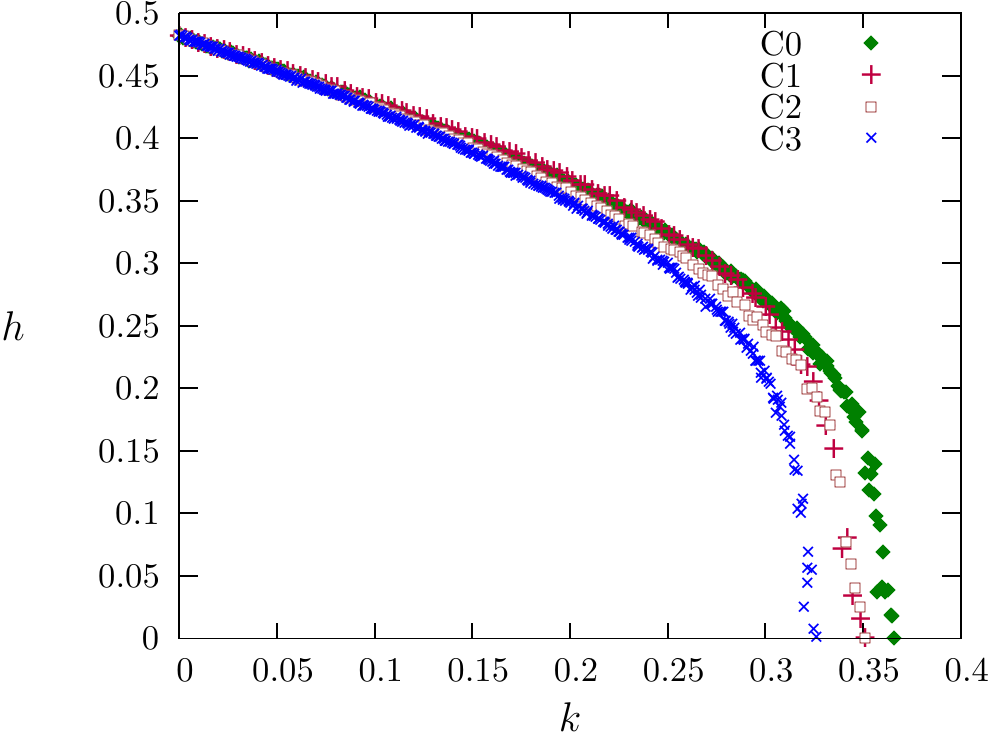} &
  \includegraphics[width=0.32\columnwidth]{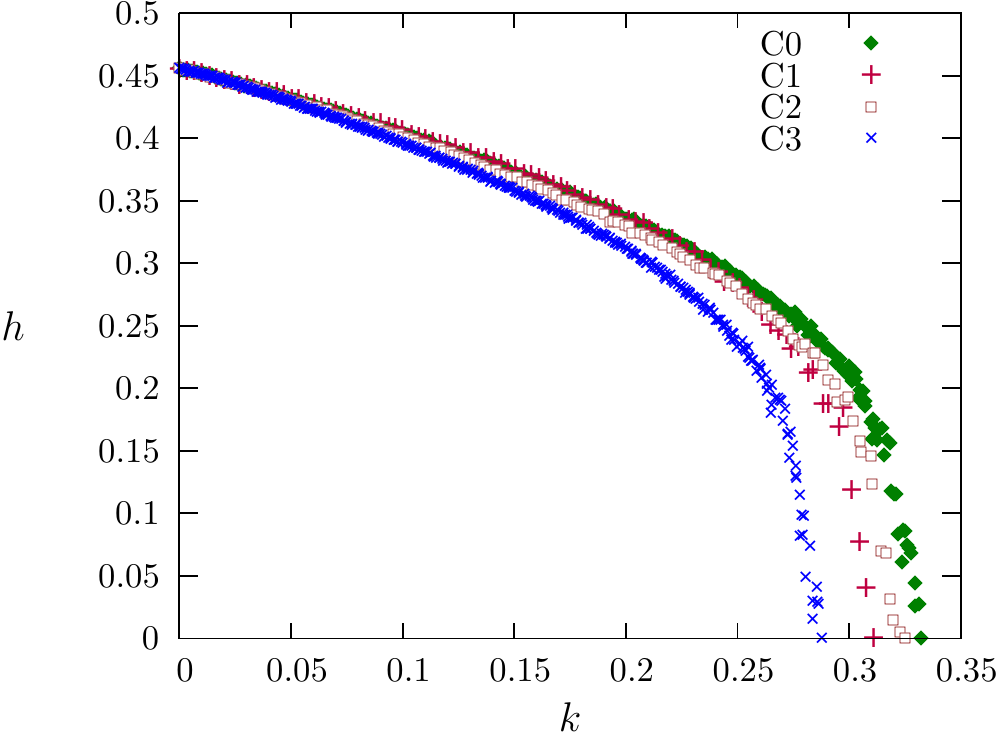} \\
   5T45&  6T28 & 7T60 \\
  \includegraphics[width=0.32\columnwidth]{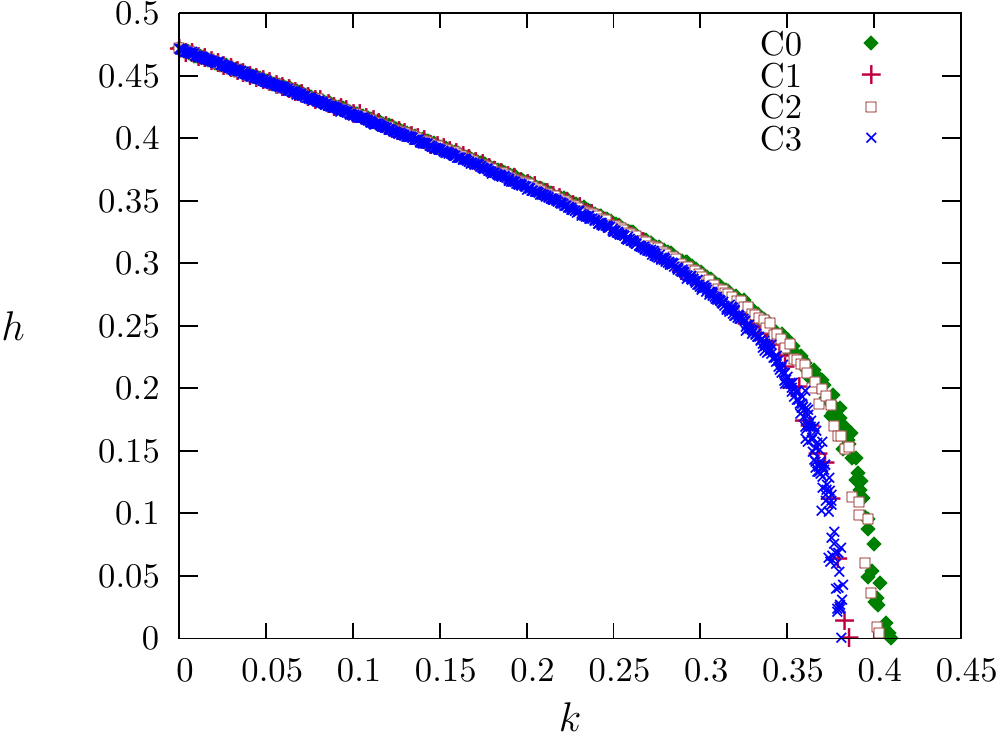} &
  \includegraphics[width=0.32\columnwidth]{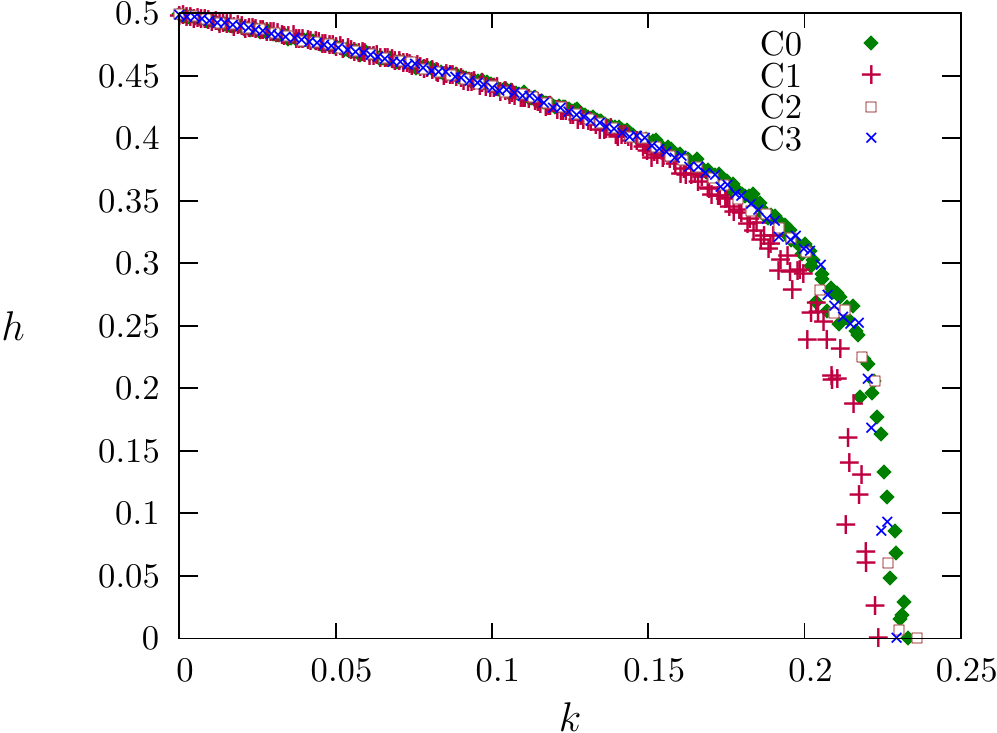} &
  \includegraphics[width=0.32\columnwidth]{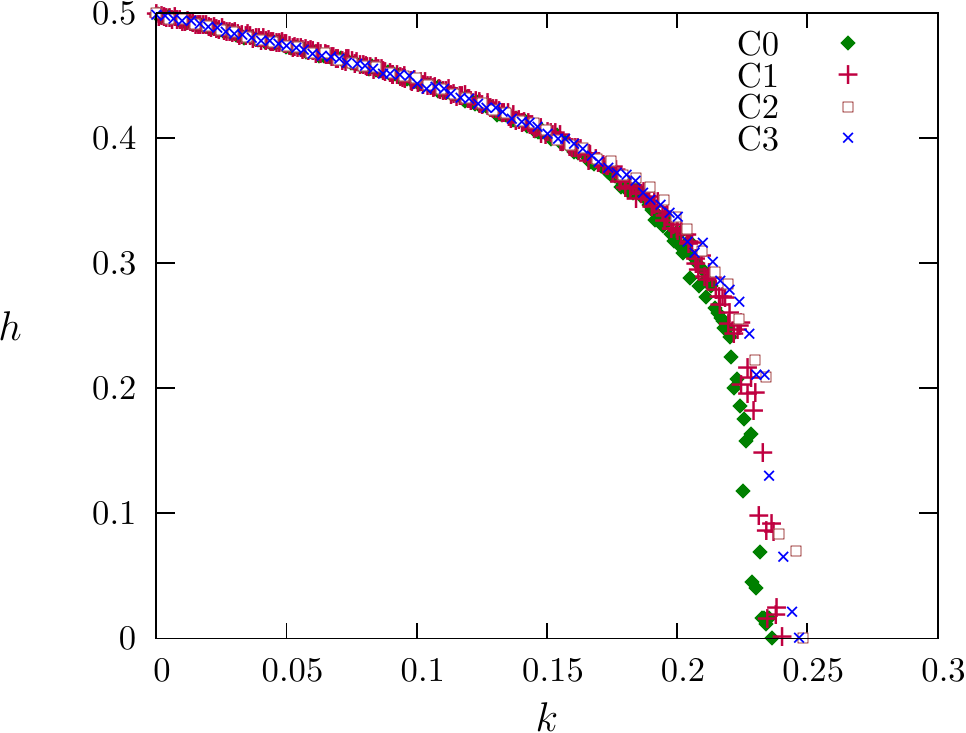}  \\
  \end{tabular}
  \caption{Phase diagram of several nonlinear CA with $N=4,000$.}
  \label{fig:nonlinear}
\end{figure}

In a large sample of nonlinear CA studied, that includes those of the previous Fig., the application of control {\bf 3} gives the smallest value of $k_c$ (which may coincide with the other types of control).  A possible explanation is that, due to the irregular behavior of the CA and the presence of an absorbing state (the synchronized or controlled state), the system will approach this absorbing state in small neighborhhoods that can grow at most linearly in time.  Thus, it is sufficient to break down the system into small portions, and for doing this it is more efficient to synchronize sites with a small (but non-zero) number of derivatives, which are presumably on the border of the non-synchronized patches, than synchronizing other sites in the bulk. Control {\bf 2} and even more control {\bf3}, preferably select sites with few non-zero derivatives, and are therefore more efficient. On the other hand, it is almost useless to synchronize sites in the bulk of the patches, with many derivatives, since they do not contribute much to the spreading of the desynchronized state.

This is not the whole story. As we show in Fig.~\ref{fig:nonlinear}, there
are CA for which control type {\bf 3} is not the best one, even if the
differences are minimal.  At present, we have no explanation for this
behavior.  


\section {Conclusions}
\label{sec:final}

We have studied the problem of master-slave synchronization and control of totalistic Boolean cellular automata via a pinching (all-or-none) mechanism. We have investigated four different cases: {\bf 0} pinching synchronization, {\bf 1} control proportional to the number of Boolean derivatives at a given site, and {\bf 2} and {\bf 3} controls for which the probability of applying the synchronization procedure diminishes with the number of Boolean derivatives. The number of nonzero local derivatives is an indicator of the expected number of defects.

For linear rules, the four types of control are equivalent and the control transition shows a crossover from a directed percolation universality class to a mean field one.  More research is needed to decide this, since It is well known that the estimation of the exponent $\beta$ requires sophisticated techniques which are left for future study.

Nonlinear CA show a counter intuitive behavior. For some of them,
control {\bf 3} is the most efficient followed by controls {\bf 2},
{\bf 0}, and {\bf 1}, although there are CA for which this is not
true.  A more exhaustive research is needed in order to find some way
to group CA according to the order in which they can be controlled.


\section*{Acknowledgements}

Interesting discussions on control of extended systems with A. El Jai are acknowledged.  Economic support for part of this work from project 25116 CONACyT-Mexico is als acknowledged.


~~

\end{document}